%% file: U1BLnsrTrh_3.tex
\documentclass[aps,prd,superscriptaddress,twocolumn,nofootinbib]{revtex4-2}

\usepackage{amsfonts,amssymb,amsmath} 
\usepackage{mathrsfs} 
\usepackage{color} 
\usepackage{graphicx}
\usepackage{dcolumn}
\usepackage{bm}

\newcommand{\C}[1]{{\mathcal #1}}

\newcommand{\BF}[1]{{\mathbf #1}} 
 
\newcommand{\nn}{\nonumber}

\newcommand{\half}{\frac 12} 
\newcommand{\third}{\frac 13} 
\newcommand{\quarter}{\frac 14} 
 
\newcommand{\sixth}{\frac 16}

\newcommand{\Slash}[1]{{\ooalign{\hfil#1\hfil\crcr\raise.167ex\hbox{/}}}}

\begin{document}


\title{Reheating consistency condition on the classically conformal \\
$U(1)_{B-L}$ Higgs inflation model}


\author{Shinsuke Kawai}
\email[]{kawai@skku.edu}
\affiliation{Department of Physics, 
Sungkyunkwan University,
Suwon 16419, Republic of Korea}
\author{Nobuchika Okada}
\email[]{okadan@ua.edu}
\affiliation{Department of Physics and Astronomy, 
University of Alabama, 
Tuscaloosa, AL35487, USA}

\date{\today}

\begin{abstract}
We revisit a cosmological scenario based on the classically conformal $U(1)_{B-L}$-extension of the Standard Model. 
Our focus is on the mechanism of reheating after inflation and the constraints on the model parameters.
In this scenario, the inflationary dynamics is driven by the $U(1)_{B-L}$ Higgs field that is nonminimally coupled to gravity and breaks the $U(1)_{B-L}$ symmetry spontaneously as it acquires a vacuum expectation value through the Coleman-Weinberg mechanism.
It is found that the reheating process proceeds stepwise, and as the decay channels of the $U(1)_{B-L}$ Higgs field are known, the reheating temperature is evaluated.
The relation between the e-folding number of inflation and the reheating temperature provides a strong consistency condition on the model parameters, and we find that the recent cosmological data gives an upper bound on the $U(1)_{B-L}$ breaking scale $v_{BL}\lesssim 10^{12}$ GeV. 
The lower bound is $v_{BL}\gtrsim 10^6$ GeV, obtained as the condition for successful reheating in this model.
The prediction for the cosmic microwave background (CMB) spectrum of this model fits extremely well with today's cosmological data.
The model can be tested and is falsifiable by near future CMB observations, including the LiteBIRD and CMB-S4. 
\end{abstract}


\maketitle


\section{Introduction}

The Standard Model of particle physics may be regarded as an outcome of gauging the global symmetries that were initially introduced for classification of particles.
The $U(1)_Y$ hypercharge originates from the work of Nakano, Nishijima \cite{Nakano:1953zz,Nishijima:1955gxk} and Gell-Mann \cite{Gell-Mann:1956iqa}, while
the $SU(2)_L$ symmetry dates back to the work of Heisenberg \cite{Heisenberg:1932dw} who introduced the concept of isospin.
The $SU(3)_c$ quantum numbers were introduced in the 1960s for the analysis of hyperons \cite{Greenberg:1964pe,Han:1965pf} (see also \cite{Tkachov:2009na}).
This view may be useful for investigating a theory beyond the Standard Model.
Indeed, there exists a $U(1)_{B-L}$ (baryon number minus lepton number) global symmetry in the Standard Model, which is usually considered accidental.
Gauging the $U(1)_{B-L}$, one obtains a theory beyond the Standard Model, that is the $U(1)_{B-L}$ gauge extended Standard Model.
It is endowed with a new $U(1)_{B-L}$ gauge boson $Z'$.
Breaking of this gauge symmetry at low energy is accomplished by a new complex scalar field $\Phi$, which plays the role of the Higgs boson for the $U(1)_{B-L}$ symmetry.
Furthermore, theoretical consistency requires three chiral fermions (right-handed neutrinos) for anomaly cancellation.
The minimal matter contents of the $U(1)_{B-L}$ extended Standard Model are thus the Standard Model particles, plus three right-handed neutrinos $N_R$ and the $U(1)_{B-L}$ Higgs boson $\Phi$, as listed in Table \ref{tab:Contents}.

The Standard Model is known to have several issues, and interestingly, many of them find natural solutions in the $U(1)_{B-L}$ extention.
The small but nonvanishing (left-handed) neutrino masses indicated by neutrino oscillations, for example, are naturally generated through the seesaw mechanism as the right-handed neutrinos acquire Majorana masses when the $U(1)_{B-L}$ symmetry is spontaneously broken.
Lepton asymmetry can also be generated by the decay of the right-handed neutrinos, which may later be converted into the baryon asymmetry of the Universe in the so-called baryogenesis via leptogenesis scenario.
Cosmic inflation may also be explained in the framework of the $U(1)_{B-L}$-extended Standard Model, as the $U(1)_{B-L}$ Higgs field $\Phi$ can play the role of the {\em inflaton}, the field responsible for the dynamics of inflation.
A simple, observationally viable and phenomenologically well-motivated model of cosmic inflation is constructed by allowing the $\Phi$ field to nonminimally couple to gravity.
The model has been a subject of much attention and has been studied actively from various aspects \cite{Iso:2009ss,Okada:2011en,Okada:2013vxa,Oda:2017zul,Okada:2019opp,Kawai:2020fjt}.

In this article, we examine the $U(1)_{B-L}$ Higgs inflation model focusing on the reheating process after inflation.
In inflationary cosmology, it is common to use the number of e-folds, $N_k$, as a parameter that quantifies the expansion of the Universe during inflation, or more specifically, between the horizon exit of the scale of the cosmic microwave background (CMB) and the end of inflation.
The typical range of $N_k$ is between 50 to 70; some uncertainly is usually assumed due to the model-dependent specifics of the reheating process.
Once the scenario of particle cosmology is specified, however, this $N_k$ is calculable in principle.
The purpose of the paper is to carry out the computation of $N_k$ in the case of the $U(1)_{B-L}$ Higgs inflation model.
The structure of the model is that the inflationary dynamics is controlled by two parameters, the breaking scale of the $U(1)_{B-L}$ symmetry $v_{BL}$ and the $U(1)_{B-L}$ gauge coupling $g_I$ at low energy.
We will find the relation between the e-folding number and those two parameters of the $U(1)_{B-L}$-extended Standard Model, and show, by solving the renormalization group equations and eliminating the uncertainties associated with the reheating process, that the prediction for the CMB spectrum is determined by $v_{BL}$ and $g_I$.
At present, those parameters are largely unconstrained, neither by collider experiments or by CMB observations.
We argue that these parameters will be severely constrained by near future precision measurements of the CMB spectrum, or the $U(1)_{B-L}$ Higgs inflation scenario will be ruled out entirely.

The rest of the paper is organized as follows.
We review the $U(1)_{B-L}$ Higgs inflation model in next section and examine the reheating process of this model in Sec.~\ref{sec:Reheat}.
In Sec.~\ref{sec:bounds} we discuss the ranges of the model parameters $v_{BL}$ and $g_I$ that are of interest to us.
We solve the cosmological evolution together with the renormalization group (RG) equations in Sec.~\ref{sec:CMB} to find the prediction of the inflationary scenario. 
We conclude in Sec.~\ref{sec:Final} with brief comments.
The appendices contain supplementary mathematical details and technical notes.

\section{Higgs inflation in the $U(1)_{B-L}$-extended Standard Model\label{sec:U1BL}}

\begin{table}[t]
\begin{tabular}{l|cccc}
  &~~~$SU(3)_c$&~~~$SU(2)_L$&~~~$U(1)_Y$~~~&~~~$U(1)_{B-L}$\\
  \hline\\
  $q_L^i$&${\BF 3}$&${\BF 2}$&$\frac 16$&$\third$\\\\
  $u_R^i$&${\BF 3}$&${\BF 1}$&$\frac 23$&$\third$\\\\
  $d_R^i$&${\BF 3}$&${\BF 1}$&$-\third$&$\third$\\\\
  $\ell_L^i$&${\BF 1}$&${\BF 2}$&$-\half$&$-1$\\\\
  $e_R^i$&${\BF 1}$&${\BF 1}$&$-1$&$-1$\\\\
  $N_R^i$&${\BF 1}$&${\BF 1}$&$0$&$-1$\\\\
  $H$&${\BF 1}$&${\BF 2}$&$\half$&$0$\\\\
  $\Phi$&${\BF 1}$&${\BF 1}$&$0$&$2$\\\\
  \hline
\end{tabular}
\caption{Representations and charges of the particle contents in the $U(1)_{B-L}$-extended Standard Model.
The subscripts $L/R$ are the chiralities and the index $i=1,2,3$ indicates the generations of the fermions.}
\label{tab:Contents}
\end{table}

We consider the minimal $U(1)_{B-L}$ extension of the Standard Model, with the gauge group $SU(3)_c\times SU(2)_L\times U(1)_Y\times U(1)_{B-L}$.
The particle contents are the Standard Model particles supplemented by three generations of singlet leptons (the right-handed neutrinos) and a complex scalar with the $U(1)_{B-L}$ charge 2, as listed in Table~\ref{tab:Contents}.
We consider the classically conformal model, that is, the scalar potential is given by
\begin{align}\label{eqn:CCV}
  V=\lambda_\Phi (\Phi^*\Phi)^2+\lambda_H (H^\dag H)^2 - \widetilde\lambda (\Phi^*\Phi)(H^\dag H),
\end{align}
where $\lambda_\Phi$, $\lambda_H$ and $\widetilde\lambda$ are dimensionless couplings.
We assume the mixing is small, $0 < \widetilde\lambda\ll 1$, so that the dynamics of $\Phi$ and $H$ are separate during inflation.
We may decompose the complex scalar $\Phi$ into two real scalars $\phi$ and $\chi$ as
\begin{align}\label{eqn:Phicomp}
  \Phi=\frac{1}{\sqrt 2}(\phi+i\chi).
\end{align}
The real component $\phi$ is assumed to have a large initial value and drive inflation, that is, it plays the role of the {\em inflaton}.
The $\chi$ field does not play any significant role below.
Including the nonminimal coupling of $\Phi$ to gravity, the Jordan frame action for the inflaton sector is 
\begin{align}
  S=\int d^4x\sqrt{-g}\left\{
  \frac{M_{\rm P}^2+\xi\phi^2}{2}R-\half(\partial\phi)^2-V_{\rm eff}\right\},
\end{align}
where $M_{\rm P}=2.435\times 10^{18}$ GeV is the reduced Planck mass, $\xi$ is a dimensionless parameter and 
\begin{align}\label{eqn:Veff}
  V_{\rm eff}=\frac{\lambda(\mu)}{4}\phi^4 + V_0
\end{align}
is the RG-improved effective action 
\cite{Coleman:1973jx,Sher:1988mj}.
The running quartic coupling $\lambda(\mu)$ is the coupling $\lambda_\Phi$ in \eqref{eqn:CCV} evaluated at the RG scale $\mu$, and the second term
\begin{align}\label{eqn:V0}
  V_0\equiv -\left.\frac{\lambda(\mu)}{4}\right|_{\phi=v_{BL}}v_{BL}^4
\end{align}
is a constant that ensures the potential vanishes at the symmetry breaking global minimum $\phi=v_{BL}$ (discussed more below).

The model is analyzed conveniently in the Einstein frame where the scalar field is minimally coupled to gravity, upon rescaling of the metric
$g_{\mu\nu}\to\Omega(\phi)\, g_{\mu\nu}$ with
\begin{align}
  \Omega(\phi)=\sqrt{1+\xi\frac{\phi^2}{M_{\rm P}^2}}.
\end{align}
The canonically normalized scalar field $\sigma$ in the Einstein frame is related to $\phi$ by
\begin{align}
  d\sigma = \frac{d\phi}{\Omega(\phi)^2}\sqrt{1+(1+6\xi)\xi\frac{\phi^2}{M_{\rm P}^2}}.
\end{align}
The scalar potential in the Einstein frame is
\begin{align}\label{eqn:VE}
  V_{\rm E}=\frac{V_{\rm eff}}{\Omega(\phi)^4}
  =\frac{\lambda(\mu)}{4\Omega(\phi)^4}\phi^4+\frac{V_0}{\Omega(\phi)^4},
\end{align}
in terms of which the slow roll parameters are defined as
\begin{align}
  \epsilon_V =& \frac{M_{\rm P}^2}{2}\left(
  \frac{V_{{\rm E},\sigma}}{V_{\rm E}}\right)^2
  =\frac{M_{\rm P}^2}{2}\left(
  \frac{V_{{\rm E},\phi}}{\sigma_{,\phi}V_{\rm E}}\right)^2,\label{eqn:epsilonV}\\
  \eta_V =& M_{\rm P}^2
  \frac{V_{{\rm E},\sigma\sigma}}{V_{\rm E}}
  =\left(\frac{M_{\rm P}}{\sigma_{,\phi}}\right)^2\left(
  \frac{V_{{\rm E},\phi\phi}}{V_{\rm E}}-\frac{\sigma_{,\phi\phi}V_{{\rm E},\phi}}{\sigma_{,\phi}V_{\rm E}}\right).\label{eqn:etaV}
\end{align}
Under the slow roll approximation, the amplitude of the curvature perturbation at comoving scale $k$ is
\begin{align}\label{eqn:As}
  P_R = \left.\frac{V_{\rm E}}{24\pi^2M_{\rm P}^4\epsilon_V}\right\vert_k,
\end{align}
which is to be compared with the measurement value\footnote{We use the
Planck 2018 TT, TE, EE + lowE + lensing central value \cite{Planck:2018jri} 
$\ln(10^{10}A_s) = 3.044$ at $k=0.05 \,\text{Mpc}^{-1}$ in the numerical computation. 
}
$A_s$ at the pivot scale $k$.
The scalar spectral index and the tensor-to-scalar ratio are expressed using the slow roll parameters as,
\begin{align}
  n_s = 1-6\epsilon_V+2\eta_V,\quad
  r = 16\epsilon_V.
\end{align}

The coupling $\lambda(\mu)$ is subject to the RG flow. 
We focus on the regime where the effects of the Yukawa couplings $y^i_M$ and the running of the nonminimal coupling $\xi$ are negligible.
Then the RG equations for the self coupling $\lambda$ and the $U(1)_{B-L}$ gauge coupling $g$ are, at 1-loop order,
\begin{align}\label{eqn:RGElambda}
  \beta_\lambda\equiv\frac{d\lambda}{d\ln\mu}=&\frac{20\lambda^2
  +96 g^4-48\lambda g^2}{16\pi^2},\\
\label{eqn:RGEg}
  \beta_g\equiv\frac{dg}{d\ln\mu}=&\frac{12 g^3}{16\pi^2}.
%
\end{align}
We interpret the quantum corrections in the presence of nonminimal coupling as follows \cite{George:2013iia}. 
The renormalization scale of the Jordan frame, in which the theory is defined, is given by the field $\phi$.
The renormalization scale (of mass dimension one) is then appropriately rescaled in the Einstein frame, in which measurements are made.
Thus the renormalization scale $\mu$ that appears in the effective potential in the Einstein frame \eqref{eqn:VE} takes the form \cite{Okada:2015zfa}
\begin{align}\label{eqn:mu}
  \mu=\frac{\phi}{\Omega(\phi)}=\frac{M_{\rm P}\phi}{\sqrt{M_{\rm P}^2+\xi\phi^2}}.
\end{align}
This is also the renormalization scale $\mu$ used in the RG equations \eqref{eqn:RGElambda} and \eqref{eqn:RGEg}.
Note that the scale $\mu$ asymptote to a constant value
$\mu\to M_{\rm P}/\sqrt\xi$ at large $\phi$, thus the RG running slows down and stops at high energy.
This behavior is in accord with the presumed UV finiteness of the theory near the Planck scale; there the metric and hence the length scale is blurred by the quantum gravity effects and above certain energy the concept of scale loses its meaning.

At low energy, the $U(1)_{B-L}$ gauge symmetry is broken by the Coleman-Weinberg mechanism.
The symmetry breaking vacuum $\phi=v_{BL}$ (where we live) satisfies the stationarity condition
\begin{align}\label{eqn:stationarity}
  \left.\frac{dV_{\rm E}}{d\phi}\right|_{\phi=v_{BL}}=0.
\end{align}
We suppose that the symmetry breaking scale is much lower than the inflationary scale ($v_{BL}\ll M_{\rm P}/\sqrt\xi$).
Then the renormalization scale is $\mu\approx\phi$ near $\phi=v_{BL}$, see \eqref{eqn:mu},
thus the distinction between the Einstein frame and the Jordan frame is unimportant at low energy.
The condition \eqref{eqn:stationarity} gives a relation between $\lambda$ and $g$,
\begin{align}\label{eqn:lambdaI}
  \lambda_I\simeq -\quarter\frac{96}{16\pi^2}g_I^4,
\end{align}
where we have used the fact that in the perturbative regime the $96\,g^4$ term dominates the right hand side of \eqref{eqn:RGElambda}.
The subscript $I$ (for IR) denotes values at the potential minimum $\phi=v_{BL}$.
Note that $\lambda_I$ is negative, as it should in the symmetry breaking minimum.
The mass of the $Z'$ boson and that of the inflaton are
\begin{align}
  m_{Z'} &=\, 2 g_I\,v_{BL},\label{eqn:Zpmas}\\
  m_\phi &=\, \left.\sqrt{\frac{d^2V_{\rm E}}{d\phi^2}}\right|_{\phi=v_{BL}}
  \simeq \frac{\sqrt 6}{\pi} g_I^2v_{BL}
  = \sqrt\frac 32\frac{g_I\,m_{Z'}}{\pi}.\label{eqn:phimass}
\end{align}
The masses of the right-handed neutrinos are given by the Majorana Yukawa coupling $y^i_{\rm M}$ as
\begin{align}
  m_{N^i_R}= \frac{y^i_{M}}{\sqrt 2}v_{BL}.\label{eqn:NRmass}
\end{align}
The offset term of the potential \eqref{eqn:V0} is now written as
\begin{align}
  V_0=\frac{3}{8\pi^2}g_I^4v_{BL}^4=\frac{3}{128\pi^2}m_{Z'}^4.
\end{align}

The symmetry breaking scale $v_{BL}$, the gauge coupling $g_I$ and the Yukawa coupling $y^i_{\rm M}$ at the potential minimum $\phi=v_{BL}$ are treated as input parameters of the model.
In particular, $v_{BL}$ and $g_I$ control the inflationary dynamics.
Let the renormalization scale at the potential minimum $\mu_I\equiv\mu(\phi=v_{BL})\approx v_{BL}$.
From there the RG equation \eqref{eqn:RGEg} for the gauge coupling is solved as,
\begin{align}\label{eqn:gsol}
  g(\mu)=\frac{g_I}{\sqrt{1-\frac{3g_I^2}{2\pi^2}\ln\frac{\mu}{\mu_I}}},
\end{align}
up to the scale $\mu$ relevant for the inflationary dynamics.
Using \eqref{eqn:gsol}, the RG equation for the self coupling \eqref{eqn:RGElambda} can be numerically integrated so that the effective potential of the inflaton \eqref{eqn:VE} can be evaluated.
The slow roll parameters are then given by \eqref{eqn:epsilonV}, \eqref{eqn:etaV} as functions of $\phi$.
To find the field value $\phi=\phi_{\rm e}$ at which inflation ends, we use the condition that one of the slow roll parameters becomes unity, $\epsilon_V(\phi_{\rm e})=1$.
The horizon exit of the CMB scale takes place at a larger value of the inflaton field $\phi=\phi_k$, and there the amplitude of the curvature perturbation \eqref{eqn:As} at the pivot scale $k$ is normalized by the observational value \cite{Planck:2018jri}.
This normalization fixes the nonminimal coupling $\xi$.
The number of e-folds for the cosmic expansion between the horizon exit of the CMB scale and the end of inflation is
\begin{align}\label{eqn:NkI}
  N_k=\frac{1}{M_{\rm P}}\int_{\phi_{\rm e}}^{\phi_k}
  \frac{d\phi}{\sqrt{2\epsilon_V(\phi)}}
  \left(\frac{d\sigma}{d\phi}\right).
\end{align}
In the standard slow roll paradigm of inflationary cosmology, it is a common practice to consider this e-folding number $N_k$ as a free parameter reflecting the uncertainty of the reheating process.
In the next section we examine the concrete reheating process of the inflationary model based on the $U(1)_{B-L}$-extended Standard Model and evaluate the e-folding number.

\section{Reheating after classically conformal $U(1)_{B-L}$ Higgs inflation\label{sec:Reheat}}

A salient feature of this cosmological model based on the classically conformal potential \eqref{eqn:CCV} is that the quartic term dominates the potential at high energy, as the mass term is generated by the Coleman-Weinberg mechanism only at the scale where the $U(1)_{B-L}$ symmetry is broken.
Thus, at the end of inflation when the amplitude of $\phi$ is still large, the potential is essentially quartic.
The symmetry breaking mass term becomes important as the oscillating amplitude of $\phi$ becomes small due to redshift.
The reheating process thus proceeds stepwise:
after inflation, the inflaton oscillates in the potential which is approximately quartic, and as the amplitude of the oscillations is damped by the redshift the inflaton starts to feel the presence of the mass term \eqref{eqn:phimass}, and then starts to oscillate in the approximately quadratic potential about the symmetry breaking minimum $\phi=v_{BL}$.
Eventually, as the Hubble expansion rate $H$ becomes comparable to the decay rate $\Gamma$ of the inflaton, the energy deposited in the inflaton is converted into the radiation of relativistic Standard Model particles and the Universe becomes thermalized.

The transition from the oscillations in the quartic-like potential to the oscillations in the quadratic-like potential is important, since the expansion rate of the Universe changes there and the prediction of the inflationary model is affected. 
At the transition, the inflaton that was swinging with a large amplitude fails to go over the central maximum of the double well potential.
This situation is characterized by the condition that the kinetic term of the inflaton becomes comparable to the potential hight at the central maximum $V_E(\phi=0)=V_0$.
Thus the inflaton energy density at this moment is approximately
\begin{align}\label{eqn:rhostar}
  \rho_\star\simeq V_0 = \frac{3}{128\,\pi^2}m_{Z'}^4.
\end{align}
We assume that the decay of the inflaton and the ensuing thermalization of the Universe takes place after this quartic-quadratic transition.
This condition is written 
\begin{align}\label{eqn:rhostarcond}
  \Gamma\lesssim H_\star,
\end{align}
with $H_\star$ the Hubble expansion rate at the transition from the quartic oscillation regime to the quadratic oscillation regime.
If the decay rate $\Gamma$ is larger than $H_\star$, the inflaton will decay immediately after the transition and thus corresponds to the case when the condition \eqref{eqn:rhostarcond} is saturated.
One may also consider possible 
decay of the inflaton condensate into radiation during the oscillations in the quartic potential \cite{Garcia:2020wiy}.
We discuss this effect in Appendix \ref{sec:meff}.
It is found that this effect is negligible if a condition slightly weaker than \eqref{eqn:rhostarcond} is satisfied.

Using the Friedman equation and \eqref{eqn:rhostar}, the condition \eqref{eqn:rhostarcond} is rewritten, up to a factor of ${\mathcal O}(1)$, as
\begin{align}
  \Gamma\lesssim\frac{m_{Z'}^2}{8\pi M_{\rm P}}.
\end{align}
We will see how this condition constrains the model parameters in Sec.~\ref{sec:bounds}.

\subsection{The number of e-folds\label{sec:Ne}}

We now evaluate the number of e-folds based on this picture, assuming otherwise the standard thermal history of the Universe.
We denote the comoving wave number of the CMB scale by $k$.
Then the scale factor $a_k$ and the Hubble parameter $H_k$ at the horizon exit of the CMB scale are related by $k=a_k H_k$.
We write the scale factor at the end of inflation as $a_{\rm e}$, 
at the quartic-quadratic transition as $a_\star$, 
at the thermalization of the Universe (end of reheating) as $a_{\rm th}$, 
at the matter-radiation equality as $a_{\rm eq}$, 
and the scale factor today as $a_0$.
Then one obtains an obvious relation
\begin{align}
  \frac{k}{a_0H_0}=\frac{a_kH_k}{a_0H_0}
  =\frac{a_k}{a_{\rm e}}\frac{a_{\rm e}}{a_\star}\frac{a_\star}{a_{\rm th}}
  \frac{a_{\rm th}}{a_{\rm eq}}\frac{a_{\rm eq}}{a_0}\frac{H_k}{H_0},
\end{align}
where 
$ H_0 = 100\,h\, {\rm km}\,{\rm s}^{-1}\,{\rm Mpc}^{-1}$ with $h=0.674$ \cite{Planck:2018vyg} is the Hubble parameter today.
The logarithm of the first factor $N_k\equiv\ln(a_k/a_{\rm e})$ is the e-folding number of inflation that we wish to evaluate.
From the end of inflation to the quartic-quadratic transition, we may write
$a_{\rm e}/a_\star=(\rho_\star/\rho_{\rm e})^{1/4}$, 
where $\rho_\star$ is \eqref{eqn:rhostar} and $\rho_{\rm e}$ is the energy density at the end of inflation, which is roughly twice the potential energy, $\rho_{\rm e}\simeq 2V_{\rm e}$.
We used the fact that when a scalar field oscillates in a quartic potential the Universe undergoes a radiation-dominant like expansion.
Likewise, from the quartic-quadratic transition to the thermalization of the Universe we may write 
$a_\star/a_{\rm th}=(\rho_{\rm th}/\rho_\star)^{1/3}$,
where $\rho_{\rm th}$ is the energy density at thermalization and we have used the fact that when a scalar field oscillates in a quadratic potential the Universe undergoes a matter-dominant like expansion.
The evaluation of the remaining factors is standard, e.g. \cite{Liddle:2003as,Martin:2010kz}.
From the thermalization to the matter-radiation equality, entropy conservation and the Stefan-Boltzmann law give
$a_{\rm th}/a_{\rm eq}=(\rho_{\rm eq}/\rho_{\rm th})^{1/4}(g_*^{\rm eq}/g_*^{\rm th})^{1/12}$, where $\rho_{\rm eq}$ is the energy density at the matter-radiation equality and $g_*^{\rm th}$, $g_*^{\rm eq}$ are the numbers of relativistic degrees of freedom at the thermalization and the matter-radiation equality, respectively.
The factor $a_{\rm eq}/a_0 = 1/(1+z_{\rm eq})$ is the redshift of the matter-radiation equality.
We use the slow roll Friedman equation to write the Hubble parameter at the time of the horizon exit of the wave number $k$ in terms of the the potential $V_k$, as 
$H_k=\sqrt{V_k/(3 M_{\rm P}^2)}$.
Assembling all those pieces we find the $e$-folding number $N_k$ between the horizon exit of the comoving wave number $k$ and the end of inflation,
\begin{align}\label{eqn:Nk}
  N_k &\equiv \ln\frac{a_{\rm e}}{a_k}=66.5-\ln h-\ln\frac{k}{a_0H_0}
+\frac{1}{12}\ln\frac{\rho_{\rm th}}{\rho_\star}\crcr
&+\frac 14\ln\frac{V_k}{2V_{\rm e}}
+\frac 14\ln\frac{V_k}{M_{\rm P}^4}
+\frac{1}{12}\left(\ln g_*^{\rm eq}-\ln g_*^{\rm th}\right).
\end{align}
Apart from the uncertainly of the reheating temperature $T_{\rm R}$ hidden in 
$\rho_{\rm th}=\pi^2 g_*^{\rm th} T_{\rm R}^4/30$, 
the e-folding number is determined by the potential \eqref{eqn:VE} and can be evaluated once the dynamics of the $U(1)_{B-L}$ Higgs field is known\footnote{
Evaluation of $V_k$ requires the value of $N_k$ \eqref{eqn:NkI} which needs to match \eqref{eqn:Nk}. This can be done consistently in numerics.
}.
In order to evaluate the reheating temperature we need to consider the decay modes of the inflaton.

\subsection{Decay of the inflaton \label{sec:decay}}

Eq.~\eqref{eqn:phimass} shows that the $Z'$ mass is heavier than the inflaton mass in the perturbative regime ($g\lesssim 1$).
Thus the decay of the inflaton into the $Z'$ boson is kinematically forbidden\footnote{
It has been pointed out in \cite{Ema:2016dny} that for $\lambda\gtrsim 3\times 10^{-4}$, violent preheating into the longitudinal mode of the gauge boson may take place in the first few oscillations of the inflaton, due to spike-like features of the conformal factor $\Omega$.
This potentially leads to an issue of unitarity as the decay products have extremely high momenta $\sim\sqrt\lambda M_{\rm P}$.
In our model this unitarity bound corresponds to $g_I\lesssim 0.13$, which is somewhat stronger than the bound from perturbativity (see \eqref{eqn:PertCond2} and Fig.~\ref{fig:ParamRegions} below).
}.
Also, the inflaton is a Standard Model singlet and it cannot decay through the Standard Model gauge interactions.
Thus the dominant decay channel of the inflaton is through the Standard Model Higgs field.

Let us use the unitary gauge
\begin{align}
  H=\begin{pmatrix}
    0\\h/\sqrt 2
  \end{pmatrix}
\end{align}
and rewrite the scalar potential \eqref{eqn:CCV} as
\begin{align}\label{eqn:V}
  V=\frac{\lambda(\mu)}{4}\phi^4+\frac{\lambda_H}{4}h^4
  -\frac{\widetilde\lambda}{4}\phi^2 h^2+V_0.
\end{align}
We may neglect\footnote{
For example, $\lambda_H\simeq 0.1$ and its quantum corrections are ${\C O}(g_2^4/16\pi^2)$, which is negligible.
}
quantum corrections for $\lambda_H$ and $\widetilde\lambda$.
The stationarity conditions 
$\partial V/\partial h=0$ and $\partial V/\partial\phi=0$ 
at the $U(1)_{B-L}$ symmetry breaking vacuum 
$h=v_H=246$ GeV and $\phi=v_{BL}$
yield
\begin{align}
  &\widetilde\lambda= 2\lambda_H
  \left(\frac{v_H}{v_{BL}}\right)^2,\label{eqn:dVdh}\\
  &\beta_\lambda+4\lambda-2\widetilde\lambda
  \left(\frac{v_H}{v_{BL}}\right)^2=0\label{eqn:dVdphi}.
\end{align}
Using \eqref{eqn:dVdh}, the last term of \eqref{eqn:dVdphi} is shown to be negligible, justifying the relation \eqref{eqn:lambdaI} that we used as the boundary conditions for the inflationary model.
We also find
\begin{align}\label{eqn:mh2}
  m_h^2=&\left.\frac{\partial^2 V}{\partial h^2}
  \right|_{\stackrel{\scriptstyle h=v_H}{\phi=v_{BL}}}
  =2\lambda_H v_H^2=\widetilde\lambda v_{BL}^2,\\
  \widetilde m^2=&\left.\frac{\partial^2 V}{\partial h\partial\phi}\right|_{\stackrel{\scriptstyle h=v_H}{\phi=v_{BL}}}=-\widetilde\lambda v_Hv_{BL}=-m_h^2\frac{v_H}{v_{BL}},\\
  m_\phi^2=&\left.\frac{\partial^2 V}{\partial\phi^2}\right|_{\stackrel{\scriptstyle h=v_H}{\phi=v_{BL}}}\simeq\frac{3g^2}{2\pi^2}m_{Z'}^2.
\end{align}
The Higgs mass is $m_h=125.25$ GeV \cite{ParticleDataGroup:2022pth}.
Now we may think of two separate cases, (i) when the inflaton mass is heavier than twice the Higgs mass $m_\phi > 2m_h$, and (ii) when the inflaton mass is lighter than twice the Higgs mass $m_\phi < 2m_h$.
Let us discuss those two cases in turn.
Below in this section we consider the fields shifted about the minimum
$h\to h+v_H$, $\phi\to\phi+v_{BL}$.

\subsubsection{$m_\phi > 2m_h$}

In this case the inflaton may decay into two Higgs through the direct coupling in \eqref{eqn:V},
\begin{align}
  \frac{\widetilde\lambda}{2}\,v_{BL}\,\phi\, h^2
  \subset\frac{\widetilde\lambda}{4}(\phi+v_{BL})^2 (h+v_H)^2
  \subset V.
\end{align}
The decay rate is
\begin{align}\label{eqn:Gammaphi}
  \Gamma_\phi=4\,\frac{(\half\widetilde\lambda\, v_{BL})^2}{8\pi m_\phi}
  =\frac{m_h^4}{8\pi m_\phi v_{BL}^2},
\end{align}
where the factor of 4 is to take into account the effects of mass mixing \cite{Cornwall:1974km,Vayonakis:1976vz,Lee:1977yc,Lee:1977eg}.
According to the standard perturbative picture of reheating\footnote{
We ignore possible nonlinear effects
\cite{Dolgov:1989us,Traschen:1990sw,Battefeld:2008bu,Kawai:2015lja}
for the sake of concreteness.
There are recent studies that suggest the perturbative picture is sufficient for typical examples \cite{Garcia:2020wiy}.
},
the inflaton starts to decay when the Hubble parameter becomes smaller than the decay rate $\Gamma_\phi$. Assuming that the thermalization is instantaneous\footnote{
Although the completion of thermalization and the start of radiation dominance are not exactly the same, the distinction is insignificant in our evaluation of \eqref{eqn:Nk}.
}, we have
$\Gamma_\phi\simeq H_{\rm th}$
and $\rho_{\rm th}$ in \eqref{eqn:Nk} is evaluated as
\begin{align}
  \rho_{\rm th}=3M_{\rm P}^2\Gamma_\phi^2.
\end{align}
The reheating temperature is then found to be
\begin{align}\label{eqn:TR}
  T_{\rm R}\simeq\left(\frac{90}{\pi^2 g_*}\right)^\quarter\sqrt{M_{\rm P}\Gamma_\phi}.
\end{align}

\begin{figure*}
\includegraphics[width=85mm]{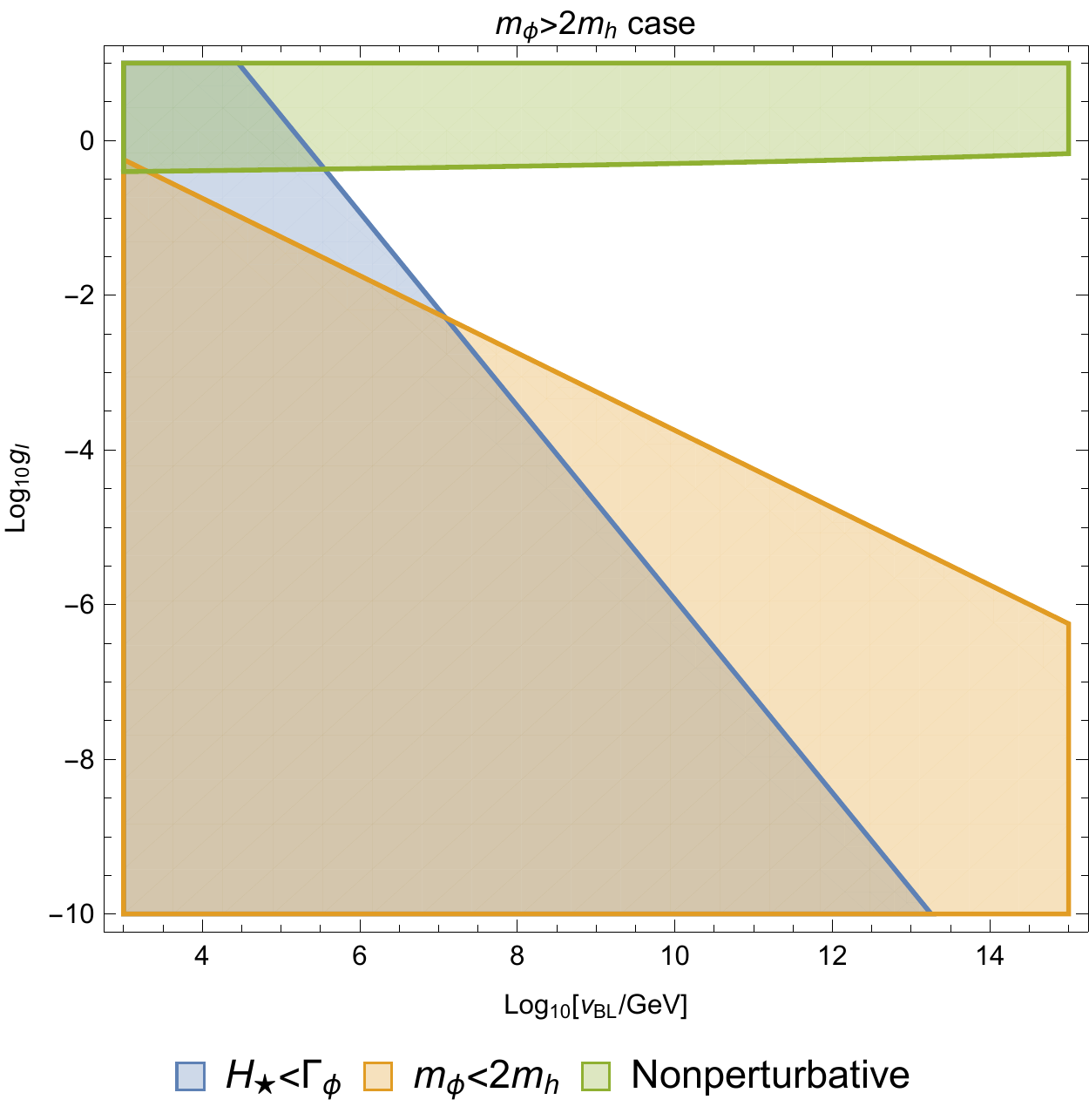}%
\includegraphics[width=85mm]{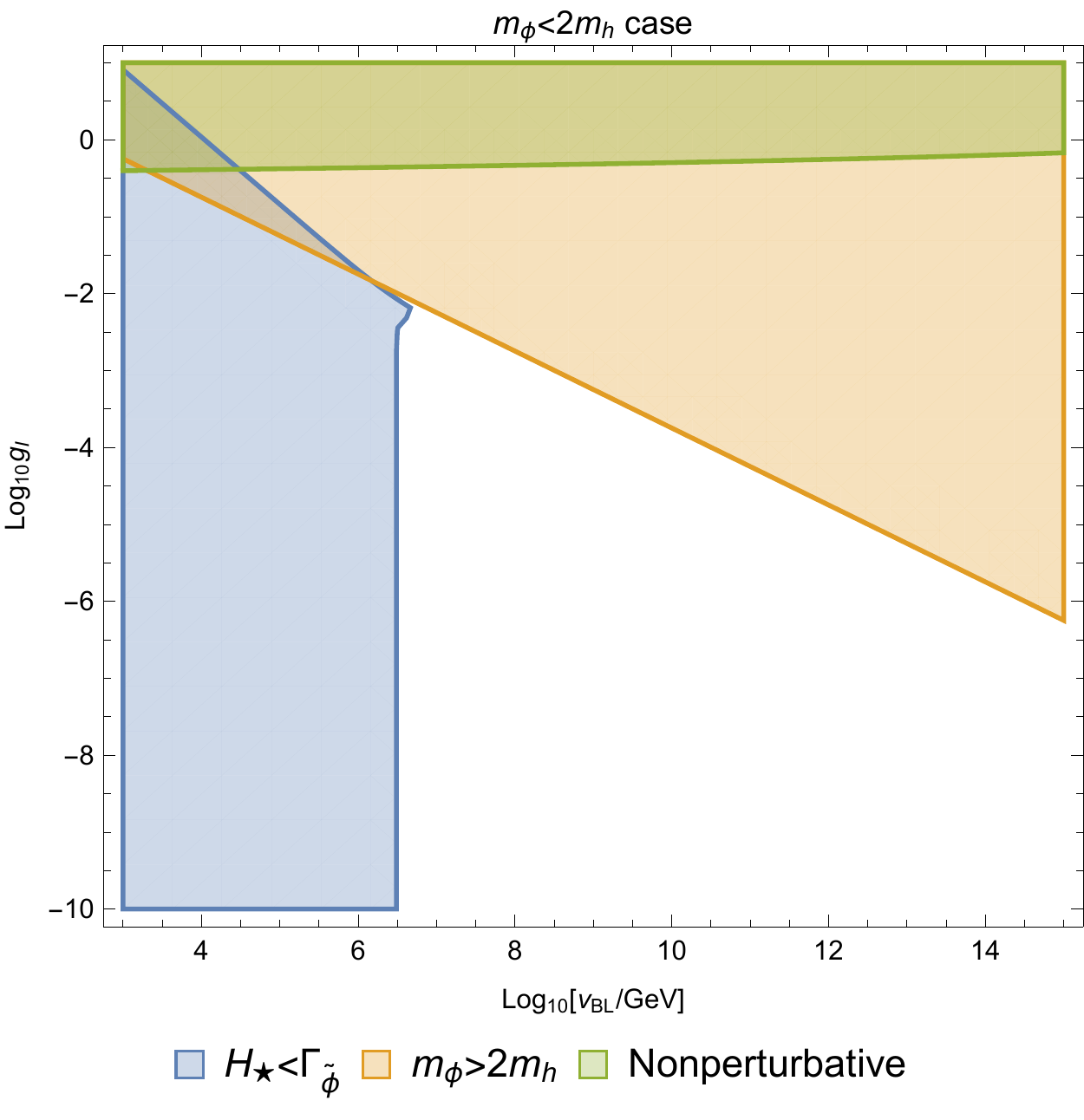}%
\caption
{\label{fig:ParamRegions}
The constraints on the parameters $(v_{BL},g_I)$, by the conditions that 
(i) the inflaton decays after it starts to oscillate about the minimum $\phi=v_{BL}$,
(ii) the inflaton is heavier (left panel) or lighter (right panel) than $2 m_h$, and 
(iii) the $U(1)_{B-L}$ gauge coupling is perturbative up until the scale of inflation.
The blank regions are unconstrained.
}
\end{figure*}

\subsubsection{$m_\phi < 2m_h$}

In the second case, when the inflaton is lighter than $2m_h$, the process $\phi\to hh$ is kinematically forbidden.
If the mass range is $m_h<m_\phi<2m_h$, the process $\phi\phi\to hh$ is possible, but reheating through this process is not possible as the decay rate $\Gamma(\phi\phi\to hh)$ redshifts faster than the Hubble expansion rate. 
When $m_\phi < 2m_h$, the inflaton may instead decay through the mixing with the Higgs boson. 
The mass matrix of the scalars
\begin{align}
  \begin{pmatrix}
    m_h^2 & \widetilde m^2\\
    \widetilde m^2 & m_\phi^2
  \end{pmatrix}
\end{align}
is diagonalized by rotating the fields
\begin{align}
  \begin{pmatrix}
    h\\\phi
  \end{pmatrix}
  =\begin{pmatrix}
    \cos\theta & \sin\theta\\
    -\sin\theta & \cos\theta
  \end{pmatrix}
  \begin{pmatrix}
    \widetilde h\\\widetilde\phi
  \end{pmatrix}.
\end{align}
The rotation angle is 
\begin{align}\label{eqn:tan2th}
  \tan 2\theta =\frac{2\widetilde m^2}{m_\phi^2-m_h^2}
  =\frac{2m_h^2}{m_h^2-m_\phi^2}\frac{v_H}{v_{BL}},
\end{align}
which is small in general, apart from the accidental narrow region of $m_\phi\simeq m_h$.
Thus the field $\widetilde h$ is almost $h$, and $\widetilde\phi$ is almost $\phi$ in generic cases.
This almost-inflaton $\widetilde\phi$ couple to the $b\overline b$, $c\overline c$, $\tau\overline\tau$ of the Standard Model with the Yukawa couplings
$y_b\sin\theta/\sqrt 2$,
$y_c\sin\theta/\sqrt 2$,
$y_\tau\sin\theta/\sqrt 2$
respectively, where $y_b$, $y_c$, $y_\tau$ are the Standard Model Yukawa couplings for $b$, $c$ and $\tau$.
The decay rate of $\widetilde\phi$ into the Standard Model particles is then
\begin{align}\label{eqn:Gammaphitilde}
  \Gamma_{\widetilde\phi}&= \frac{m_\phi}{8\pi}
  \left(3\frac{m_b^2}{v_H^2}+3\frac{m_c^2}{v_H^2}+\frac{m_\tau^2}{v_H^2}\right)\sin^2\theta\crcr
  &\simeq 4.0\times 10^{-5} m_\phi\sin^2\theta,
\end{align}
where we have used 
$m_b = 4.2$ GeV, $m_c = 1.3$ GeV and $m_\tau = 1.8$ GeV.
Thus, using
$2\sin^2\theta=1-1/\sqrt{1+\tan^22\theta}$ and \eqref{eqn:tan2th}
the decay rate is determined by $m_\phi$ and $v_{BL}$.
The condition for the decay is
$\Gamma_{\widetilde\phi}\simeq H_{\rm th}$.
Thus the Friedman equation gives 
$\rho_{\rm th}\simeq 3M_{\rm P}^2\Gamma_{\widetilde\phi}^2$
and the reheating temperature is similar to \eqref{eqn:TR}, with 
$\Gamma_{\phi}$ now replaced by $\Gamma_{\widetilde\phi}$.

\section{Constraints on the parameters\label{sec:bounds}}

Before discussing the cosmological prediction of the inflationary model in the next section, let us summarize the constraints on the two parameters $v_{BL}$ and $g_I$.

First of all, we assume that perturbative quantum field theory is valid up to the scale of inflation.
As the condition of perturbativity we demand that the gauge coupling is perturbative up to the Planck scale\footnote{
It may be somewhat more natural to consider
$\alpha\equiv g^2/4\pi<1$ at $\mu\to M_{\rm P}/\sqrt\xi$
as the criterion of perturbativity.
We however use the slightly tighter condition \eqref{eqn:PertCond1} for the sake of practical convenience, as it is $\xi$-independent and leaves some margin from the singular regions that are numerically difficult to handle.
}
\begin{align}\label{eqn:PertCond1}
  g(\mu=M_{\rm P})<1.
\end{align}
Using the solution \eqref{eqn:gsol} of the RG equation this condition is written, with
$\mu_I=\mu(\phi=v_{BL})\simeq v_{BL}$,
\begin{align}\label{eqn:PertCond2}
  g_I<\frac{1}{\sqrt{1+\frac{3}{2\pi^2}\ln\frac{M_{\rm P}}{\mu_I}}}.
\end{align}

Other conditions concern the decay of the inflaton, so let us consider the two separate cases as we did in the previous section.

\begin{enumerate}
  \item $m_\phi>2m_h$\\
Using \eqref{eqn:phimass}, the condition $m_\phi>2m_h$ is written
\begin{align}\label{eqn:2mhcond}
  g_{\rm I}\gtrsim\left(
  \frac{2\pi m_h}{\sqrt 6\,v_{BL}}\right)^\half.
\end{align}
The condition \eqref{eqn:rhostarcond} that the inflaton decay after the quartic-quadratic transition is written using \eqref{eqn:Gammaphi} as 
\begin{align}\label{eqn:rhostarcond1}
  g_{\rm I}\gtrsim\left(
  \frac{\pi M_{\rm P}m_h^4}{4\sqrt 6\, v_{BL}^5}\right)^\quarter.
\end{align}
It can be checked that \eqref{eqn:2mhcond} is a stronger constraint than \eqref{eqn:rhostarcond1} when
$v_{BL}>1.23\times 10^7$ GeV.

  \item $m_\phi<2m_h$\\
In this case the condition $m_\phi<2m_h$ is written
\begin{align}\label{eqn:2mhcond2}
  g_{\rm I}\lesssim\left(
  \frac{2\pi m_h}{\sqrt 6\,v_{BL}}\right)^\half,
\end{align}
and the condition that the inflaton decays after the transition \eqref{eqn:rhostarcond} reads
\begin{align}\label{eqn:rhostarcond2}
  \Gamma_{\widetilde\phi}\lesssim\frac{m_{Z'}^2}{8\pi M_{\rm P}},
\end{align}
where $\Gamma_{\widetilde\phi}$ is evaluated using 
\eqref{eqn:tan2th} and \eqref{eqn:Gammaphitilde}.
\end{enumerate}

Fig.~\ref{fig:ParamRegions} shows the constraints on the symmetry breaking scale $v_{BL}$ and the gauge coupling at low energy $g_I$ as described above.
The green region is excluded by the perturbativity condition \eqref{eqn:PertCond2} and the blue region is excluded by the requirement that the decay of the inflaton takes place after the quartic-quadratic transition, \eqref{eqn:rhostarcond1} for the left panel and \eqref{eqn:rhostarcond2} for the right panel.
The orange region is excluded by the condition on the inflaton mass, $m_\phi>2\,m_h$ for the left panel and $m_\phi<2\,m_h$ for the right panel.
Light inflaton $m_\phi\lesssim 10^6$ TeV is excluded in both cases, and there are both upper and lower bounds for $g_I$ in the $m_\phi>2\,m_h$ case, whereas in the $m_\phi<2\,m_h$ case $g_I$ is only bounded from above.

\begin{figure}
\includegraphics[width=85mm]{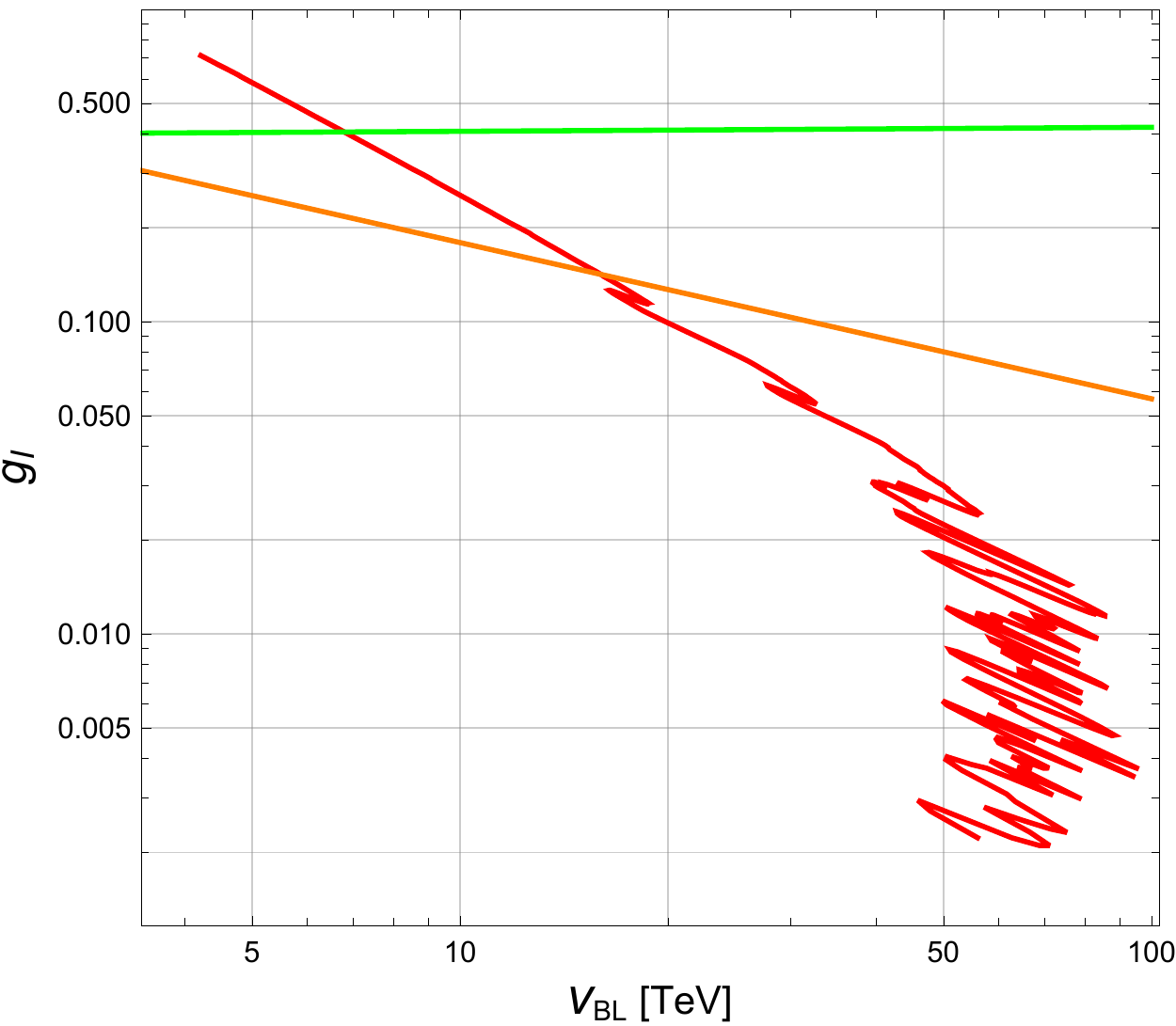}%
\caption
{\label{fig:ATLAS}
Constraints from the ATLAS experiments \cite{ATLAS:2019erb} at Run 2 of the Large Hadron Collider (the center-of-mass energy $\sqrt s=13$ TeV and integrated luminosity 139 fb${}^{-1}$), recast into bounds on the parameters $v_{BL}$ and $g_I$ of the $U(1)_{B-L}$-extended Standard Model.
The region lower left to the red curve is excluded. 
The green and orange lines are respectively the perturbativity limit and the $m_\phi=2m_h$ line as in Fig.~\ref{fig:ParamRegions}.
}
\end{figure}

Let us also comment on the bounds that come from the collider experiments. 
Fig.~\ref{fig:ATLAS} shows the bounds on $v_{BL}$ and $g_I$ obtained from the search for high-mass dilepton resonance by the ATLAS detector in the Large Hadron Collier \cite{ATLAS:2019erb} (139 fb${}^{-1}$ proton-proton collisions at a center of mass energy $\sqrt s=13$ TeV).
The lower left region to the red curve is excluded.
Also shown are the green and orange lines, that are the perturbativity bound and the $m_\phi=2m_h$ line as in Fig.~\ref{fig:ParamRegions}.
Comparing Fig.~\ref{fig:ATLAS} and Fig.~\ref{fig:ParamRegions}, the bounds from the ATLAS experiments are seen to provide no further constraints on the parameter space of the inflationary model as the region is already excluded by the condition $H_\star<\Gamma_{\phi}$ (in the case of $m_\phi>2\,m_h$) or $H_\star<\Gamma_{\widetilde\phi}$ (in the case of $m_\phi<2\,m_h$).

\section{CMB spectrum of the $U(1)_{B-L}$ Higgs inflation model\label{sec:CMB}}

\begin{figure*}
\includegraphics[width=100mm]{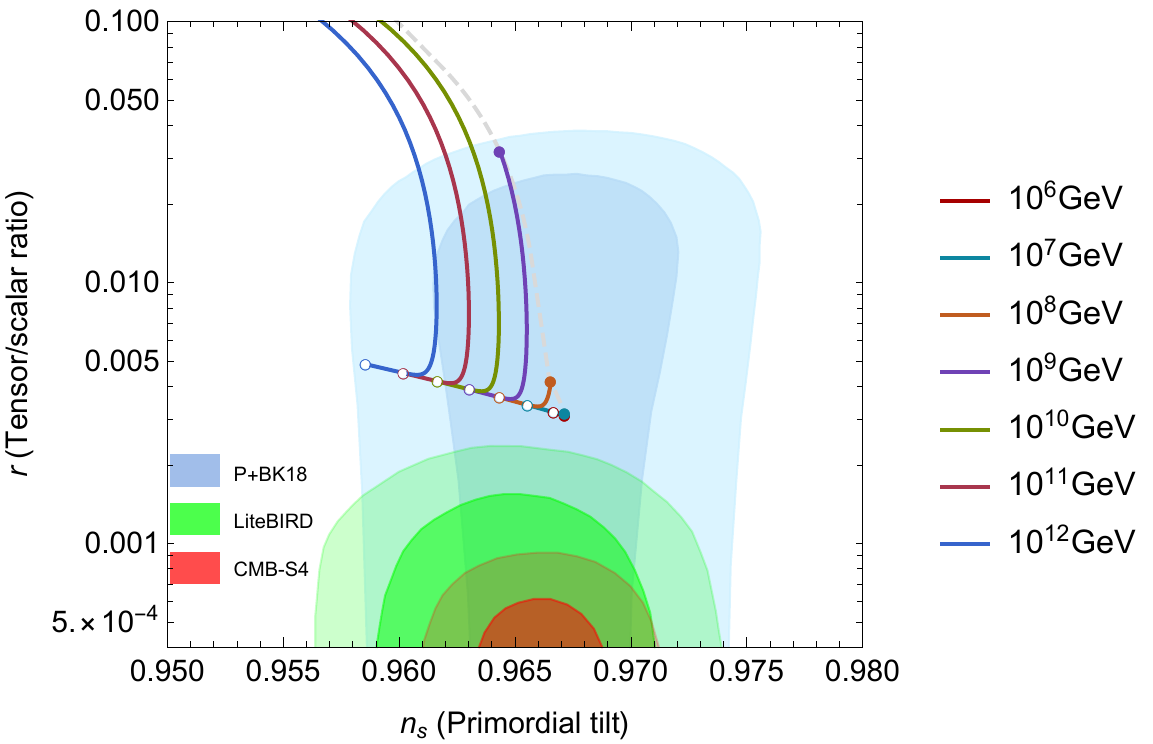}%
\includegraphics[width=79mm]{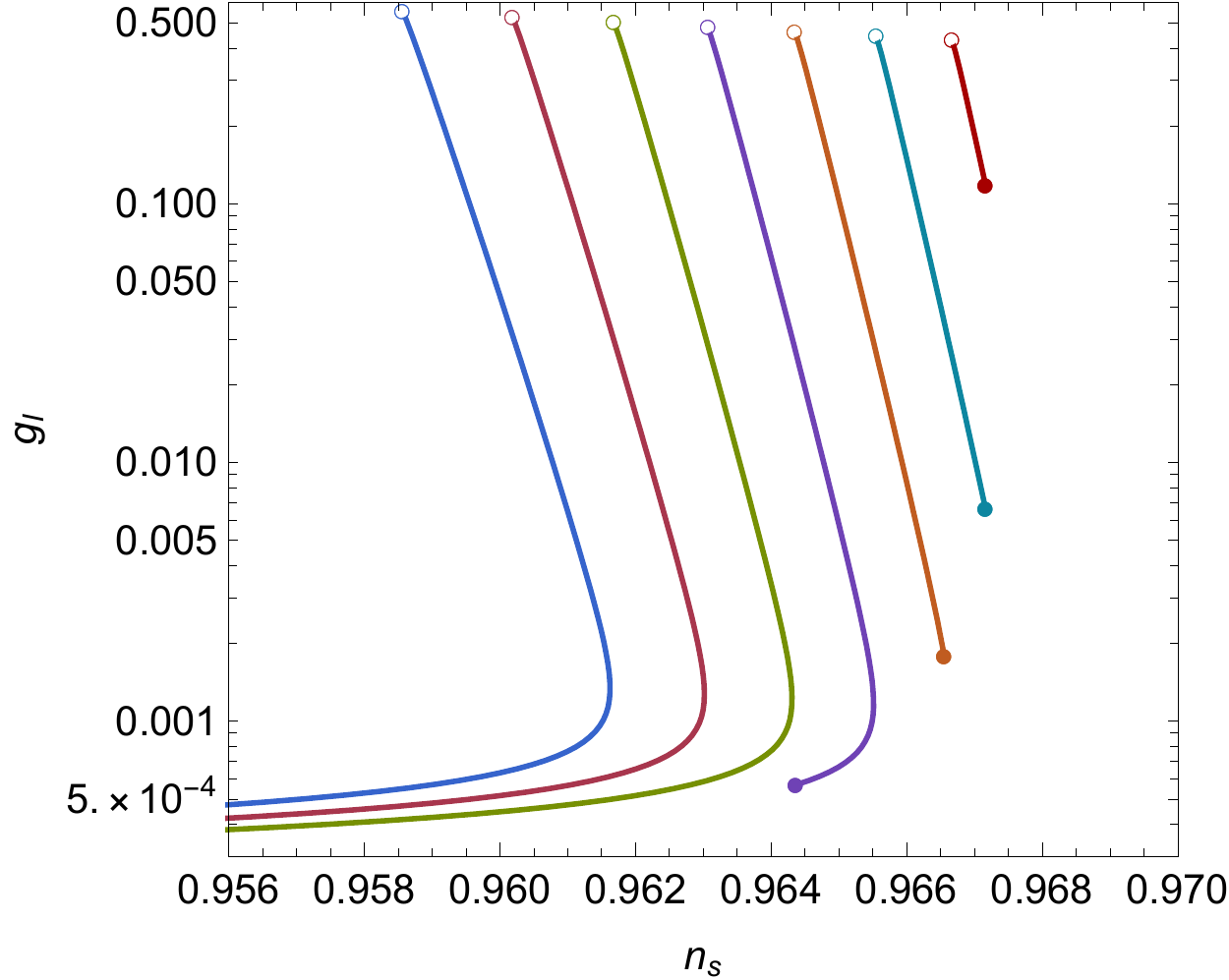}%
\caption{
\label{fig:nsrandgI}
The prediction of the $U(1)_{B-L}$ Higgs inflation model, with the requirement of the reheating consistency taken into account.
This is the $m_\phi>2\,m_h$ case of Fig.~\ref{fig:ParamRegions}.
The value of $v_{BL}$ is varied as $10^6$, $10^7$, $10^8$, $10^9$, $10^{10}$, $10^{11}$, $10^{12}$ GeV.
The left panel shows the scalar spectral index $n_s$ and the tensor-to-scalar ratio $r$.
The right panel shows $n_s$ and the $U(1)_{B-L}$ coupling $g$ at the symmetry breaking minimum $\phi=v_{BL}$.
The end points marked with $\bullet$ correspond to the lower bound of $g_I$, limited by the $m_\phi>2m_h$ condition or the $\Gamma<H_\star$ condition.
The end points mark with $\circ$ correspond to the upper bound of $g_I$ given by the perturbativity condition.
The background contours are the 68\% and 95\% confidence level
Planck+BICEP/Keck 2018 results
\cite{BICEP:2021xfz} (blue), and the LiteBIRD \cite{LiteBIRD:2022cnt} (green) and CMB-S4 \cite{Abazajian:2019eic} (red) 1- and 2-$\sigma$ prospects for a fiducial model with $r=0$.
}
\end{figure*}

Let us now discuss the prediction of the cosmological model.

\subsection{Numerical method\label{sec:procedure}}

In order to determine the set of parameters that meet the consistency requirements and to calculate the resulting spectrum of the CMB, we employ the following procedure to solve the slow roll equation of motion and the RG equations.
For a specified value of the symmetry breaking scale $v_{BL}$, we choose a set of parameters $N_k^{(\rm test)}$ and $g_I$ so that $(v_{BL}, g_I)$ is within the allowed parameter region discussed in Sec.~\ref{sec:bounds}. 
Then the slow roll equation and the RG equations can be numerically integrated, using $N_k^{(\rm test)}$ as the e-folding number of \eqref{eqn:NkI} to identify the field value $\phi_k$ at the horizon exit of the CMB scale.
The value of the nonminimal coupling $\xi$ is adjusted so the the amplitude of the curvature perturbation $P_R$ matches the Planck normalization value \cite{Planck:2018jri} at the pivot scale.
Then the cosmological evolution is determined by the set of three parameters $(v_{BL},N_k^{(\rm test)},g_I)$, and we may evaluate the e-folding number defined by the formula \eqref{eqn:Nk}, which, in general, differs from the value of $N_k^{(\rm test)}$.
We then make a scan of the parameter $g_I$ (but $v_{BL}$ and $N_k^{(\rm test)}$ kept fixed) to see if $N_k$ of \eqref{eqn:Nk} can be adjusted to be the same value as $N_k^{(\rm test)}$.
If $g_I$ satisfying $N_k=N_k^{(\rm test)}$ is found in the range of Sec.~\ref{sec:bounds}, then the solution meets all consistency requirements.
If this procedure fails, then that means there is no cosmological solution compatible with the reheating consistency requirement.

We carried out the parameter scan within the allowed regions of Fig.~\ref{fig:ParamRegions}, and have found solutions satisfying these requirements.
In the case of $m_\phi>2m_h$, for $v_{BL}\gtrsim 10^6$ GeV there exist consistent cosmological solutions between the upper and lower bounds of $g_I$. 
In contrast, when $m_\phi<2m_h$ we only found consistent solutions in narrow vicinities of $m_\phi=m_h$, where the mixing angle between $\phi$ and $h$ becomes $\theta=\pi/4$.
While this situation may be of some phenomenological interest, it is outside of our initial assumption that the inflaton dynamics is independent of the Standard Model Higgs field during inflation, and we thus will not examine this case further.

\subsection{Numerical results\label{sec:Plots}}

We thus discuss the results for the $m_\phi>2\,m_h$ case below.
Fig.~\ref{fig:nsrandgI} shows the solutions.
The left panel is the prediction for the CMB spectrum, the primordial tilt $n_s$ against the tensor-to-scalar ratio $r$ for the consistent solutions as described above. 
The curves indicate solutions for fixed values of $v_{BL}=10^6$ GeV to $10^{12}$ GeV and the background contours shaded in blue are the 68\% and 95\% confidence level constraints of the recent Planck+BICEP/Keck 2018 constraints \cite{BICEP:2021xfz}.
It is seen that the prediction of the model comfortably sits inside the 68\% contour, up to $v_{BL}\lesssim 10^{12}$ GeV.
The right panel shows the same set of solutions on the $n_s$-$g_I$ plane. 
In both panels, the endpoints marked with a filled/blank circle correspond to the lower/upper bound of $g_I$ shown on the left panel of Fig.~\ref{fig:ParamRegions}.
The $v_{BL}=10^6$ GeV solution is seen to be trimmed by the $\Gamma_\phi<H_\star$ constraint, as one can see by comparing with Fig.~\ref{fig:ParamRegions}, left panel.

In Fig.~\ref{fig:nsrandgI}, the prospect constraint contours by the LiteBIRD and CMB-S4, for a $r=0$ fiducial model are shown in green and red. 
The prediction of the cosmological model studied here is clearly outside the 2-$\sigma$ contours, and thus would be strongly disfavored if those projects bring null results. 
If, on the other hand, the tensor mode is detected, the measurements of the CMB spectrum would give significant constraints on the parameter space of the $U(1)_{B-L}$ Higgs inflation model.

\section{Final remarks\label{sec:Final}}

We have examined the reheating process of the inflationary scenario based on the $U(1)_{B-L}$ extension of the Standard Model, and formulated the condition of consistency in terms of the number of e-folds.
We then solved the equation of the inflationary dynamics along with the RG equations to identify solutions that meet these requirements.
The results show that the predictions of the CMB spectrum are in excellent agreement with current observational constraints.
It is also suggested that the proposed model could be tested by future experiments, such as LiteBIRD and CMB-S4.
Our aim was to address the previously overlooked aspects of model construction and to provide a clearer prediction for cosmological observables by incorporating the consistency condition from the reheating process.

The primary focus of this paper has been the analysis of a simple inflationary model, which is characterized by two key parameters: the $U(1)_{B-L}$ breaking scale ($v_{BL}$) and the $U(1)_{B-L}$ gauge coupling ($g_I$) at low energy. 
The $U(1)_{B-L}$ extension of the Standard Model is a well-motivated theory beyond the Standard Model, and this example may be considered as one of the best candidate cosmological scenarios based on particle phenomenology.
Clearly, our analysis can be extended to more involved cosmological models.
For instance, the $U(1)_{B-L}$ model can be extended to the $U(1)_X$ model that allows for the mixing of the $U(1)_{B-L}$ and $U(1)_Y$ gauge symmetries without violating the anomaly cancellation condition, as described for example in \cite{Kawai:2020fjt}.
Additionally, inflationary models based on supersymmetric extensions of the Standard Model, such as those discussed in \cite{Arai:2013vaa,Kawai:2021hvs}, may also be worthy of exploration.
As upcoming observational cosmology projects are poised to bring new results in the near future, particularly with regards to the CMB B-model polarization, it is a promising time to re-evaluate the reheating dynamics of these inflationary scenarios.


\begin{acknowledgments}
%
This work was supported in part by the National Research Foundation of Korea Grant-in-Aid for Scientific Research Grant No. NRF-2022R1F1A1076172 (SK) and by the United States Department of Energy Grant No. DE-SC0012447 (N.O.).
\end{acknowledgments}


\appendix

\section{Evaluation of the slow roll parameters\label{sec:SRparams}}

In our computation, the slow roll parameters \eqref{eqn:epsilonV}, \eqref{eqn:etaV} are used to identify the field value at the end of inflation, to find the normalized amplitude of the curvature perturbation, as well as to evaluate the spectrum of the CMB.
The expressions of \eqref{eqn:epsilonV} and \eqref{eqn:etaV} involve $V_{\rm E}$ \eqref{eqn:VE} as well as its $\phi$-derivatives.
The concrete expressions of the first and second derivatives of $V_{\rm E}$ employed in our analysis are obtained using the RG equations \eqref{eqn:RGElambda}, \eqref{eqn:RGEg} and the relation \eqref{eqn:mu} for $\mu(\phi)$.
These are
\begin{align}
  V_{{\rm E},\phi}&\equiv \frac{dV_{\rm E}}{d\phi}\crcr
  &=\frac{\phi}{(1+\xi\phi^2)^3}\left\{\left(
  \lambda+\frac{\beta_\lambda}{4}
  \right)\phi^2-4\xi V_0\right\},
\end{align}
\begin{align}
  V_{{\rm E},\phi\phi}&\equiv\frac{d^2V_{\rm E}}{d\phi^2}\crcr
  &=\frac{1}{(1+\xi\phi^2)^4}\Bigg\{
  3\lambda\phi^2(1-\xi\phi^2)
  +\frac{\beta_\lambda}{4}\phi^2(7-3\xi\phi^2)\crcr
  &+\frac{\lambda\phi^2(25\lambda^2-90\lambda g^2+156g^4)}{32\pi^4}
  -4\xi V_0 \left(1-5\xi\phi^2\right)
  \Bigg\}.\nn\\
\end{align}
We have set the reduced Planck mass to unity, $M_{\rm P}=1$.
It is then straightforward to find the expressions of the slow roll parameters \eqref{eqn:epsilonV}, \eqref{eqn:etaV} as functions of the field $\phi$.

We also used
\begin{align}
  \sigma_{,\phi}\equiv &\frac{d\sigma}{d\phi}=
  \frac{\sqrt{1+(1+6\xi)\xi\phi^2}}{1+\xi\phi^2},\\
  \sigma_{,\phi\phi}\equiv &\frac{d^2\sigma}{d\phi^2}=
  -\frac{\xi\phi}{(1+\xi\phi^2)^2}
  \frac{1-6\xi+(1+6\xi)\xi\phi^2}{\sqrt{1+(1+6\xi)\xi\phi ^2}}.
\end{align}

\section{Inflaton decay during oscillations in the quartic potential \label{sec:meff}}

In the main text we did not consider the decay of the inflaton when it is oscillating in the quartic potential.
As discussed e.g. in \cite{Garcia:2020wiy}, the oscillating inflaton may be interpreted to form a {\em condensate} obtaining its mass from the averaged periodic motions, and decay into radiation during this regime.
Here we discuss this effect, first evaluating the criteria for which the decay can be efficient, and then give an alternative picture of it based on particle scattering.

\subsection{Efficiency of the energy depletion}

We consider the classically conformal effective action \eqref{eqn:CCV} of the $U(1)_{B-L}$ Higgs inflation model
\begin{align}
  V=\lambda |\Phi|^4+\lambda_H (H^\dag H)^2 
  -\widetilde\lambda |\Phi|^2(H^\dag H)+(\text{1-loop}),
\end{align}
and decompose $\Phi$ in the unitary gauge into a lowly varying background field $\phi_o$ and the field $\phi$ on that background,
\begin{align}\label{eqn:phio}
  \Phi=\frac{1}{\sqrt 2}(\phi_o(t)+\phi).
\end{align}

In this regime, the field $\phi$ has a time-dependent effective mass
\begin{align}\label{eqn:meff2}
  m_{\rm eff}^2\equiv \frac{d^2 V}{d\phi^2}\Bigg|_{\phi_o} 
  = 3\lambda\phi_o^2,
\end{align}
and the coupling between $\phi$ and $H^\dagger H$ is given by
\begin{align}
  {\C L}_{\rm int}=\widetilde\lambda\phi_o\phi H^\dagger H.
\end{align}
The decay amplitude for $\phi\to H^\dagger H$ is
\begin{align}
  \sum_{\rm spins}\left|{\C M}\right|^2 = 2\widetilde\lambda^2\phi_o^2,
\end{align}
and thus the decay width is found to be time dependent,
\begin{align}\label{eqn:Gammaphicdst}
  \Gamma(t) 
  = \frac{\widetilde{\lambda}^2\phi_o^2}{8\pi m_{\rm eff}}
  = \frac{\sqrt 3}{24\pi}\frac{\widetilde\lambda^2}{\sqrt\lambda}\left|\phi_o(t)\right|.
\end{align}
As the universe expands like radiation dominated $a\propto\sqrt t$ in this regime, the inflaton redshifts as
$\phi_o\sim \phi_{\rm e}\sqrt{t_{\rm e}/t}$,
where $t_{\rm e}$ and $\phi_{\rm e}$ are the cosmic time and the background inflaton value at the end of inflation.
Using $H(t=t_{\rm e})=1/2t_{\rm e}$ and the slow roll equation of motion, we find
\begin{align}\label{eqn:te}
  t_{\rm e} \sim \sqrt\frac{3}{\lambda}\frac{M_{\rm P}}{\phi_{\rm e}^2}.
\end{align}
The decay width \eqref{eqn:Gammaphicdst} is then written 
\begin{align}
  \Gamma(t) = \frac{\sqrt 3}{24\pi}\frac{\widetilde\lambda^2}{\sqrt\lambda}\phi_{\rm e}\sqrt\frac{t_{\rm e}}{t}
  \equiv \Gamma_0\sqrt\frac{t_{\rm e}}{t}.
\end{align}

The energy density of the inflaton $\rho_\phi$ and that of the radiation $\rho_{\rm rad}$ evolve according to
\begin{align}\label{eqn:Boltzrhophi}
  &\frac{d\rho_\phi}{dt}+4H\rho_\phi+\Gamma(t)\rho_\phi = 0,\\
  &\frac{d\rho_{\rm rad}}{dt}+4H\rho_{\rm rad}-\Gamma(t)\rho_\phi = 0,
\end{align}
where $H=1/2t$.
The total energy density $\rho_{\rm total}=\rho_\phi+\rho_{\rm rad}$ thus evolves as $\rho_{\rm total}\propto a^{-4}$.
The equation \eqref{eqn:Boltzrhophi} is solved as
\begin{align}
  \rho_\phi(t)=\rho_\phi(t_{\rm e})\left(\frac{t_{\rm e}}{t}\right)^2 \exp\left\{-2\Gamma_0 t_{\rm e}\left(\sqrt\frac{t}{t_{\rm e}}-1\right)\right\}.
\end{align}
The factor $(t_{\rm e}/t)^2$ is due to the dilution by the cosmic expansion and the exponential factor with
\begin{align}
  D(t)\equiv 2\Gamma_0 t_{\rm e}\left(\sqrt\frac{t}{t_{\rm e}}-1\right)\simeq 2\Gamma_0 t_{\rm e}\sqrt\frac{t}{t_{\rm e}}
\end{align}
represents the energy transmission into radiation.

Thus the depletion of the inflaton energy by the decay into radiation is negligible if
\begin{align}\label{eqn:cdstdecay}
  D(t=t_\star)\lesssim 1,
\end{align}
where $t_\star$ is the time when the inflaton starts to oscillate in the quadratic potential. 
Using \eqref{eqn:te} and evaluating $\phi_o\sim v_{BL}$ at $t=t_\star$, the condition \eqref{eqn:cdstdecay} is equivalent to
\begin{align}\label{eqn:cdstdecay2}
  \frac{1}{4\pi}\frac{\widetilde\lambda^2}{\lambda}
  \frac{M_{\rm P}}{v_{BL}}\lesssim 1.
\end{align}
Now using \eqref{eqn:dVdh}, \eqref{eqn:mh2} and evaluating $\lambda$ in our model of the Coleman-Weinberg symmetry breaking as
\begin{align}
  \lambda\simeq\lambda_{\rm eff}
  \equiv\sixth\frac{d^4V_{\rm E}}{d\phi^4}\Bigg|_{\phi=v_{BL}}
  =\frac{11}{\pi^2}g_I^4,
\end{align}
the condition \eqref{eqn:cdstdecay2} gives a lower bound on $g_I$:
\begin{align}
  g_I\gtrsim\left(\frac{\pi}{44}\right)^\quarter
  \left(\frac{m_h}{v_{BL}}\right)
  \left(\frac{M_{\rm P}}{v_{BL}}\right)^\quarter.
\end{align}
This is seen to be a slightly weaker condition than \eqref{eqn:rhostarcond1}.

\subsection{Scattering picture}

Instead of the decay $\phi\to H^\dagger H$, one may alternatively consider the scattering process $\phi\phi\to H^\dagger H$.
The initial $\phi$'s are assumed to be condensed, at rest with energy $\omega_\phi$.
The number density of the inflaton quanta
$\phi$ must then obey the Boltzmann equation
\begin{align}\label{eqn:Boltzmann}
  \frac{dn_\phi}{dt}+3Hn_\phi
  = - (\sigma v_{\rm rel})n_\phi^2,
\end{align}
with
\begin{align}\label{eqn:sigmavrel}
  \sigma v_{\rm rel}\sim\frac{\widetilde\lambda^2}{16\pi\omega_\phi^2}.
\end{align}
The energy of the inflaton quanta $\omega_\phi$ may be evaluated as
\begin{align}
  \omega_\phi^2\simeq 
  m_{\rm eff}^2 = \frac{d^2V}{d\phi^2}\Bigg|_{\phi_o}
  =3\lambda\phi_o^2.
\end{align}
We shall show that \eqref{eqn:Boltzmann} is equivalent to \eqref{eqn:Boltzrhophi}, up to a numerical factor.

We first note that $n_\phi=\rho_\phi/\omega_\phi$, and that $\omega_\phi$ redshifts as $\omega_\phi\propto\phi_o\propto 1/a$, so that
\begin{align}
  \frac{d\rho_\phi}{dt}+4H\rho_\phi
  =&\left(\frac{dn_\phi}{dt}\omega_\phi
  +4Hn_\phi+n_\phi\frac{d\ln\omega_\phi}{dt}\right)\crcr
  =&\omega_\phi\left(\frac{dn_\phi}{dt}+3Hn_\phi\right).
\end{align}
Then \eqref{eqn:Boltzrhophi} is rewritten as
\begin{align}
  \frac{dn_\phi}{dt}+3Hn_\phi+\frac{\Gamma(t)}{n_\phi}n_\phi^2=0.
\end{align}
Now using 
$\omega_\phi\simeq\sqrt{3\lambda}\phi_o$, 
$\rho_\phi\simeq \frac 34\lambda\phi_o^4$ and the expression \eqref{eqn:Gammaphicdst} we find
\begin{align}
  \frac{\Gamma(t)}{n_\phi}
  = \frac{\Gamma(t)}{\rho_\phi}\omega_\phi
  \simeq \frac{4\Gamma(t)}{3\lambda\phi_o^4}\omega_\phi
  \simeq \frac{12\lambda\Gamma(t)}{\omega_\phi^3}
  \simeq \frac{1}{2\pi}\frac{\widetilde\lambda^2}{\omega_\phi^2},
\end{align}
coinciding with \eqref{eqn:sigmavrel} up to a factor of $\sim 8$, that may be attributed to the approximations we have used.

\input{U1BLnsrTrh_3.bbl}

\end{document}

%% file: U1BLnsrTrh_3.bbl
%